\documentclass[12pt,usenames,dvipsnames,a4paper]{article}

\usepackage{soul}

\usepackage[utf8]{inputenc}
\usepackage[T1]{fontenc}

\usepackage{changepage}
\usepackage{amsmath,amsfonts,amssymb}
\usepackage{epsf,amsmath,bbold,amsfonts,stmaryrd}
\usepackage{mathrsfs}
\usepackage{appendix}
\usepackage{caption}
\usepackage{color}
\usepackage{datetime}
\usepackage{float}
\usepackage{graphicx}
\usepackage[colorlinks,hyperindex,breaklinks]{hyperref}
\hypersetup{pageanchor=false,citecolor=red,urlcolor=red}
\usepackage{indentfirst}
\usepackage[numbers,square,comma,sort&compress,merge]{natbib} 
\usepackage{subfig}
\numberwithin{equation}{section}
\usepackage{mathtools}
\usepackage{ytableau}
\usepackage{tabu}
\usepackage{esvect}

\usepackage{calc}

\allowdisplaybreaks

\def\arctanh{\mathrm{arctanh}}

\def\c{\chi}

\def\eps{\varepsilon}
\def\f{\frac}
\def\g{\gamma}

\def\G{\Gamma}

\def\l{\left}

\def\m{\mu}
\def\n{\nu}

\def\p{\partial}

\def\r{\right}
\def\s{\sigma}
\def\t{\tau}

\def\x{\xi}

\def\z{\zeta}

\def\be{\begin{equation}}
\def\ee{\end{equation}}

\def\bea{\begin{eqnarray}}
\def\eea{\end{eqnarray}}

\def\ba{\begin{array}}
\def\ea{\end{array}}

\def\bc{\begin{center}}
\def\ec{\end{center}}

\def\bl{\begin{flushleft}}
\def\el{\end{flushleft}}

\def\br{\begin{flushright}}
\def\er{\end{flushright}}

\def\bi{\begin{itemize}}
\def\ei{\end{itemize}}

\def\bt{\begin{tabular}}
\def\et{\end{tabular}}

\makeatletter

\newsavebox\myboxA
\newsavebox\myboxB
\newlength\mylenA

\newcommand*\xoverline[2][0.75]{%
    \sbox{\myboxA}{$\m@th#2$}%
    \setbox\myboxB\null
    \ht\myboxB=\ht\myboxA%
    \dp\myboxB=\dp\myboxA%
    \wd\myboxB=#1\wd\myboxA
    \sbox\myboxB{$\m@th\overline{\copy\myboxB}$}
    \setlength\mylenA{\the\wd\myboxA}
    \addtolength\mylenA{-\the\wd\myboxB}%
    \ifdim\wd\myboxB<\wd\myboxA%
       \rlap{\hskip 0.5\mylenA\usebox\myboxB}{\usebox\myboxA}%
    \else
        \hskip -0.5\mylenA\rlap{\usebox\myboxA}
         {\hskip 0.5\mylenA\usebox\myboxB}%
    \fi}
\makeatother

\def\be{\begin{equation}}
\def\ee{\end{equation}}

\def\bea{\begin{eqnarray}}
\def\eea{\end{eqnarray}}

\def\f{\frac}

\def\p{\partial}

\usepackage{xspace}
\newcommand*{\ie}{i.e., }
\newcommand*{\eg}{e.g., }

\newcommand*{\eq}{eq.\@\xspace}

\newcommand*{\cf}{cf.\@\xspace}

\usepackage{xifthen}
\newcommand*\diff{\mathrm{d}} 
\newcommand*\ldiff[2][]{ \ifthenelse{\isempty{#1}}{ \diff
#2}{\diff^#1#2} \,} 
\let\limitint\int 
\renewcommand{\int}{\limitint \!} 

\begin{document}

\begin{titlepage}

\vspace*{-1cm}

\begin{adjustwidth}{-1.3cm}{-.7cm}

\begin{center}
	\bf \Large{Weyl-invariant Einstein-Cartan gravity: \\
		unifying the strong CP and hierarchy puzzles}
\end{center}
\end{adjustwidth}

\begin{center}
\textsc{Georgios K. Karananas,$^\star$~Mikhail 
Shaposhnikov,$^\dagger$~Sebastian Zell\,$^\ddagger$}
\end{center}

\begin{center}
\it {$^\star$Arnold Sommerfeld Center\\
Ludwig-Maximilians-Universit\"at M\"unchen\\
Theresienstra{\ss}e 37, 80333 M\"unchen, Germany\\
\vspace{.4cm}
$^\dagger$Institute of Physics \\
\'Ecole Polytechnique F\'ed\'erale de Lausanne (EPFL) \\ 
CH-1015 Lausanne, Switzerland\\
\vspace{.4cm}
$^\ddagger$Centre for Cosmology, Particle Physics and Phenomenology -- CP3,\\
Universit\'e catholique de Louvain,\\
B-1348 Louvain-la-Neuve, Belgium
}
\end{center}

\begin{center}
\small
\texttt{\small georgios.karananas@physik.uni-muenchen.de}  \\
\texttt{\small mikhail.shaposhnikov@epfl.ch} \\
\texttt{\small sebastian.zell@uclouvain.be} 
\end{center}

\begin{abstract} 

We show that the minimal Weyl-invariant Einstein-Cartan gravity in combination
with the Standard Model of particle physics contains just one extra scalar
degree of freedom (in addition to the graviton and the Standard Model fields)
with the properties of an axion-like particle which can solve the strong
CP-problem. The smallness of this particle's mass as well as of the
cosmological constant is ensured by tiny values of the gauge coupling
constants of the local Lorentz group. The tree value of the Higgs boson mass
and that of Majorana leptons (if added to the Standard Model to solve the
neutrino mass, baryogenesis and dark matter problems) are very small or
vanishing, opening the possibility of their computability in terms of the
fundamental parameters of the theory due to nonperturbative effects.

\end{abstract}

\end{titlepage}

\tableofcontents

\section{Introduction}
\label{sec:intro}

Gauge symmetries are pivotal for our understanding of Nature. In the Standard
Model (SM) of particle physics, the strong and electroweak interactions arise
from the gauging of the internal $SU(3)\times SU(2)\times U(1)$ groups.
Gravity can also be derived as gauge theory of the Poincar\'e group~\cite
{Utiyama:1956sy,Kibble:1961ba,Sciama:1962}, which singles out the
Einstein-Cartan (EC) formulation of General Relativity (GR)~\cite
{Cartan:1922,Cartan:1923,Cartan:1924,Cartan:1925} with the tetrad and
connection the associated gauge fields; confer for instance the classic
review~\cite{Hehl:1976kj}. Yet another important gauge symmetry is that of
Weyl~\cite{Weyl:1917rtf,Weyl:1918ib,Weyl:1919fi,Weyl1923-WEYRZM}, which
changes locally the length of rulers.

This paper aims to show that the minimal theory which embraces the SM and
Weyl-invariant quadratic gravity in the EC formulation has several attractive
features. It contains one extra (pseudo)scalar degree of freedom in addition
to the graviton and the particles of the SM. This new degree of freedom has
all the properties of an axion-like particle (ALP). The constructed theory
provides insights into several puzzles of the SM. One of the gauge couplings
associated with the Lorentz group determines the classical value of the
cosmological constant, and so we know from phenomenology that this parameter
has to be extremely small. It is then logical to expect that the other gauge
coupling(s) be equally small. Amazingly, those determine the Higgs and ALP
masses (the former being generated dynamically in our framework). Moreover,
since our theory automatically accommodates a coupling of the ALP to the
topological charge density of QCD, it can relax the strong CP-angle $\bar
{\theta}$ close to zero, \ie leads to a purely gravitational solution to the
strong CP-problem. Evidently, all of this fully resonates with the
expectation that gravity is the weakest force \cite
{Arkani-Hamed:2006emk}.~\emph{In short, we provide a common gravitational
origin for the seemingly independent tunings of the cosmological constant,
Higgs mass, and $\theta$-angle.}

Taken at face value, the classical theory constructed along these lines
contradicts to observations---the Higgs boson mass is too small. However,
this should not be considered as the weakness of the proposal,~\emph{this is
actually its strength}. Indeed, the classical action is not the whole story.
The smallness or absence of the Higgs and ALP masses persists in
scale-invariant quantum perturbation theory, as the zero values of these
parameters correspond to fixed points of their renormalization group
evolution~\cite{tHooft:1973mfk,Wetterich:1983bi}. This makes them in
principle computable in terms of the parameters of the underlying theory,
once nonperturbative effects are taken into account \cite
{Coleman:1973jx, Weinberg:1976pe, Linde:1977mm, Shaposhnikov:2018xkv,
Shaposhnikov:2018jag, Shaposhnikov:2020geh}. As fas as the ALP is concerned,
these effects are well understood and are associated with the non-trivial
topological structure of the QCD vacuum \cite
{Callan:1976je,Jackiw:1976pf}, leading to nonperturbative mass
generation \cite{tHooft:1976rip,tHooft:1976snw}. If the QCD contribution is
larger than the gravitationally induced tree-level mass, the strong CP
problem is solved in exact similarity with the conventional axion  \cite
{Peccei:1977hh,Weinberg:1977ma,Wilczek:1977pj}. Concerning the Higgs boson
mass, its nonperturbative origin is more speculative and requires further
investigations, which we leave for future work. A possibility certainly worth
exploring is that  the hierarchy between the Fermi and Planck scales be
attributed to the semiclassical suppression of gravitational-Higgs instanton
amplitudes, as proposed in~\cite{Shaposhnikov:2018xkv,Shaposhnikov:2020geh}
(see also~\cite{Shaposhnikov:2018jag,Karananas:2020qkp}).

 Exact or approximate scale and Weyl invariance have both been evoked
 previously in attempts to solve the hierarchy problem, see~\cite
 {Wetterich:1983bi,Bardeen:1995kv,
 Wetterich:1987fk,Wetterich:1987fm,Dehnen:1992jc,Wetterich:1994bg,Cervantes-Cota:1995ehs,Foot:2007iy,Shaposhnikov:2008xb,Shaposhnikov:2008xi,Shaposhnikov:2008ar,Shaposhnikov:2009nk,GarciaBellido:2011de,Blas:2011ac,GarciaBellido:2012zu,Bezrukov:2012hx,Monin:2013gea,Tavares:2013dga,Khoze:2013uia,Csaki:2014bua,Rubio:2014wta,Ghilencea:2015mza,Karam:2015jta,Trashorras:2016azl,
 Karananas:2016grc,Ferreira:2016vsc,Karananas:2016kyt,Karam:2016rsz,Ferreira:2016wem,Ferreira:2016kxi,Ghilencea:2016dsl,Shkerin:2016ssc,Rubio:2017gty,Tokareva:2017nng,Casas:2017wjh,Ferreira:2018itt,Ferreira:2018qss,Shaposhnikov:2018jag,Burrage:2018dvt,Lalak:2018bow,Gorbunov:2018llf,Iosifidis:2018zwo,Casas:2018fum,Shkerin:2019mmu,Herrero-Valea:2019hde,Karananas:2019fox,Rubio:2020zht,Karananas:2020qkp,Hill:2020oaj,Piani:2022gon,Karananas:2021gco}
 for a non-exhaustive list of references. An important ingredient of previous
 studies of {\em exact} symmetries is an extra dilaton field $\chi$,
 implementing the idea of spontaneously broken scale invariance. The main
 drawback of these theories is that both the Weyl and scale symmetries allow
 for a dilaton-Higgs interaction of the type $\alpha \c^2 H^\dagger H$, where
 $H$ is the Higgs field. In the presence of an ALP $a$, the allowed terms in
 the action are $c_a \c^2 a^2$. With the dilaton vacuum expectation value of
 the order of the Planck mass, phenomenology dictates that the coupling
 constants $\alpha, c_a$ have to be very small without any obvious reason why
 this should be the case. In the theory we present in this paper, there is no
 any additional scalar field, all degrees of freedom are contained in the
 gauge fields of the gauged Poincar\'e group. The ``dangerous'' dilaton type
 couplings are either absent or automatically tiny.

Similar considerations comprise~\cite{Edery:2015wha,Edery:2019txq} based on
the idea of ``restricted Weyl invariance'' introduced in~\cite
{Edery:2014nha}, the works~\cite
{Ghilencea:2018dqd,Ghilencea:2018thl,Ghilencea:2021lpa}~based on the Weyl
conformal geometry~and~\cite{Shtanov:2023lci}~using global scale invariance.
All have a gravitational sector which contains the associated scalar
curvature squared, as well as couplings to the SM via a gravitational-Higgs
portal. The usual Einstein-Hilbert action with a nonvanishing cosmological
constant is recovered in the spontaneously broken phase. The Higgs sector
also gets nontrivially modified in that a tachyonic mass is induced, so
electroweak symmetry  breaking may be traced to gravity. However, no ALP of
gravitational origin was considered in~\cite
{Edery:2015wha,Edery:2019txq,Ghilencea:2018dqd,Ghilencea:2018thl,Ghilencea:2021lpa,Shtanov:2023lci}.

It is important to point out that our approach differs significantly from the
conventional axion solution to the strong CP-problem, which is based on
postulating a Peccei-Quinn (PQ) symmetry~\cite
{Peccei:1977hh,Weinberg:1977ma,Wilczek:1977pj}. In the most popular
models~\cite{Kim:1979if,Shifman:1979if,Zhitnitsky:1980tq,Dine:1981rt}, at
least six new degrees of freedom are required in addition to the SM fields
and  graviton.\footnote{Two degrees of freedom are contained in the PQ-field.
In KSVZ~\cite{Kim:1979if,Shifman:1979if}, one complex scalar field and a new
massive quark are required, while DFSZ \cite
{Zhitnitsky:1980tq,Dine:1981rt} relies on two complex scalar fields, where
one is a doublet with respect to the $SU(2)$ weak isospin and the other one
is a singlet.}~Apart from the question about the physical origin of the
assumed PQ symmetry, one ``unnaturally'' small number---the CP-violating
angle $\bar \theta \lesssim 10^{-10}$---is replaced  by two others: the ratio
of the electroweak $v$ and the PQ $F$ scales, $\left
(v/F\right)^2 \lesssim 10^{-14}$, and the parameter measuring the ``quality''
of the symmetry, $\left(M/F\right)^2\lesssim 10^{-50}$, where $M$ is the
scale around which the Peccei-Quinn symmetry is explicitly broken. In
contrast, our model only brings into a single new degree of freedom, namely
the axion itself. Moreover, no additional tuning is introduced apart from the
smallness of the cosmological constant, which exists anyway. In summary, one
may say that we get three tunings for the price of one.
	
This observation is particularly interesting since ``quantum breaking'' has
provided indications that eternal de Sitter states must not exist in a
consistent theory of quantum gravity~\cite
{Dvali:2013eja,Dvali:2014gua,Dvali:2017eba,Dvali:2018fqu,Dvali:2018jhn,
Dvali:2020etd,Dvali:2021kxt}~(see also~\cite
{Obied:2018sgi,Andriot:2018wzk,Garg:2018reu,Ooguri:2018wrx} for similar
conjectures in string theory). However, the presence of $\bar
{\theta}$-vacua in QCD would lead to de Sitter states, and so the only way
out is to make the $\bar{\theta}$-angle unphysical by the addition of an
axion -- its existence becomes a mandatory consistency requirement,
independently of any naturalness considerations~\cite
{Dvali:2018dce,Dvali:2022fdv,Dvali:2023llt}. Our proposal completes the
picture as gravity also provides the necessary ALP.

Finally, we mention that our approach is fully compatible with the neutrino
Minimal Standard Model ($\nu$MSM)~\cite{Asaka:2005an,Asaka:2005pn}, which is
a minimal extension of the SM in the neutrino sector capable of addressing
simultaneously the experimental problems of the latter: neutrino masses and
oscillations, dark matter, baryon asymmetry of the Universe. Weyl symmetry
forces the tree-level Majorana masses of the heavy neutral leptons (HNLs) of
the $\nu$MSM to be zero, which would be incompatible with phenomenology since
successful baryogenesis cannot take place. In full analogy to the situation
with the Higgs, we can speculate that non-perturbative effects generate
masses for the HNLs. This is in line with the common lore that gravity breaks
all global symmetries (see e.g.~\cite{Kallosh:1995hi}). In this case, the
classical action would have a global $B-L$ symmetry in the absence of HNL
masses ($B$ and $L$ being the baryon and lepton numbers, respectively), and
the breaking of $B-L$ \`a la Nambu-Jona-Lasinio~\cite
{Nambu:1961tp,Nambu:1961fr} can potentially lead to Majorana masses for the
HNLs, even though the order of magnitude of this effect remains obscure and
has never been computed. As an additional bonus, the EC formulation of GR
provides a mechanism for generating the HNLs in the early Universe so that
the lightest of them can provide the observed abundance of dark matter in a
wide range of masses~\cite{Shaposhnikov:2020aen}.

The paper is organized as follows. In Section~\ref{sec:toy-model}, we define
the action of the theory, guided by the principle of having only
Weyl-invariant terms at most quadratic in curvatures and with a consistent
particle spectrum. The principle of minimality relates the Weyl gauge field
to the vector part of torsion. In Section~\ref{sec:metric_equivalent}, we
integrate out torsion to obtain the action in the metric formulation of GR
such that the presence of the dynamical pseudoscalar mode originating from
the gravitational Holst term, the ALP, is made explicit. We also fix here the
free parameters of the theory, associated with the cosmological constant and
the ALP mass. In Section~\ref{sec:strong_solution}, we show how the ALP can
indeed solve the strong CP problem. We proceed in Section~\ref
{sec:higgs_coupling}, by adding the Higgs sector. We demonstrate that the
gravitationally-generated Higgs mass is tiny and that the gravitational
solution to the strong CP-puzzle persists. In Section~\ref{sec:euclidean}, we
discuss how the theory can be analytically continued to Euclidean spacetimes
and point out that it may be possible to obtain a positive-definite action.
In Section~\ref{sec:conclusions}, we conclude. In Appendix~\ref
{app:diagonalization}, we show that a nontrivial redefinition of field
variables casts in diagonal form the kinetic sector of the scalars (in its
full generality) and with the Higgs canonical.

\textbf{Conventions}. We use the conventions of~\cite
 {Rigouzzo:2022yan,Rigouzzo:2023sbb}. Lowercase Greek and capital Latin
 letters denote spacetime and Lorentz indices, respectively. Both the
 spacetime $g_{\mu\nu}$ and Minkowski $\eta_{AB}$ metrics  have signature $
 (-1,+1,+1,+1)$. For the gamma matrices we use 
\begin{equation} 
\label{gammaConvention}
\left\{\gamma_A, \gamma_B \right\} = - 2 \eta_{AB} \ ,
 ~~~\gamma_5 = -i \gamma^0 \gamma^1 \gamma^2 \gamma^3 =
 i \gamma_0 \gamma_1 \gamma_2 \gamma_3 \ ,
\end{equation}   
and for the Levi-Civita tensor we take
\begin{equation} 
\label{epsilonConvention}
	\epsilon_{0123}=1=-\epsilon^{0123} \ ,
\end{equation}   
for spacetime and Lorentz indices.

Making the Poincar\'e group local necessitates the introduction of the gauge
fields associated with translations and Lorentz transformations. In the
standard jargon these are called tetrad $e^A_\m$ and spin connection $\omega^
{AB}_\m$. Their respective field-strength tensors are torsion
\be
\label{eq:torsion_tensor_T} 
T^A_{\m\n} = \p_\m e^A_\n -\p_\n e^A_\m + \omega^A_{\m B} e^B_\n - \omega^A_{\n B} e^B_\n \ .
\ee
and curvature
\be
\label{eq:curvature_tensor_F}
F^{AB}_{\m\n} = \p_\m \omega^{AB}_\n - \p_\n\omega^{AB}_\m + \omega^A_{\m C}\omega^{CB}_\n - \omega^A_{\n C}\omega^{CB}_\m \ ,
\ee
respectively.

The covariant derivative of a spacetime vector $A^\m$ is defined as
\begin{equation} 
\label{covariantDerivative}
\nabla_\n A^\m= \partial_\n A^\m + \Gamma^\m_{~\n \rho}A^\rho \ ,
\end{equation}
where $\Gamma^\m_{~\n\rho}= e^\m_A\l(\p_\n e^A_\rho+\omega^A_{\n B}e^B_
{\rho}\r)$ is the affine connection. The associated ``affine'' curvature
tensor $R$ reads
\be
\label{eq:curvature_tensor} 
R_{~\sigma \mu \nu}^{\rho}=\partial_{\mu} \Gamma_{~\nu \sigma}^{\rho}-
\partial_{\nu} \Gamma_{~\mu \sigma}^{\rho}+\Gamma_{~\mu \lambda}^
 {\rho} \Gamma_{~\nu \sigma}^{\lambda}-\Gamma_{~\nu \lambda}^{\rho} \Gamma_
 {\mu \sigma}^{\lambda} \ ,
\ee 
while ``affine'' torsion is simply given by the antisymmetric part of the
connection 
\begin{equation} 
\label{torsionDefinition}
T^\m_{~\n \rho} \equiv \Gamma^\m_{~\n \rho}-\Gamma^\m_{~\rho\n} \ .
\end{equation} 
For later convenience in  the computations  we decompose this into its three
irreducible  pieces: the vector 
\be 
\label{eq:torsion_vector} v^{\m}  =g_{\n\rho}T^{\n \m \rho} \ , 
\ee 
the pseudovector
\be
\label{eq:torsion_pseudovector}
a^{\m}=E^{\m \n \rho \s}T_{\n \rho \s} \ , 
\ee
and the sixteen-component reduced torsion tensor
\be
\label{eq:torsion_reduced_tensor}
\tau_{\m \n \rho}	=T_{\m \n \rho}  +\frac{1}{3}\left(g_{\m \n}v_{\rho}- g_
 {\m\rho}v_{\n} \right)-\frac{1}{6}E_{\m \n \rho \s} a^\s \ ,
\ee
that satisfies
\be
g^{\m\rho}\tau_{\m\n\rho} = E^{\m\n\rho\s}\tau_{\n\rho\s} = 0 \ .
\ee
To declutter the notation we also introduced the densitized $\epsilon$
\be
E^{\m\n\rho\s} = \f{\epsilon^{\m\n\rho\s}}{\det(e)} \ ,~~~E_{\m\n\rho\s} = \det(e)\epsilon_{\m\n\rho\s} \ ,
\ee
with $\det(e)$ the determinant of the tetrad. 

\section{The simplest scenario:~induced cosmological constant 
and axion-like particle of gravitational origin}
\label{sec:toy-model}

We are interested in the most general theory that fulfills the following
criteria (see \cite
{Karananas:2021zkl,Karananas:2021gco,Rigouzzo:2023sbb,Barker:2024dhb} for
similar approaches):
\begin{enumerate} 
\item The gravitational action is composed of terms that are at most quadratic
 in the field strength tensors $F^{AB}_{\m\n}$ and $T^A_{\m\n}$ (equivalently 
  $R_{\mu\nu\rho\sigma}$ and $T^\m_
 {~\n \rho}$). 
\item The theory is Weyl invariant at the classical level.\footnote
 {The quantum aspects of this theory, including the Weyl anomaly will be
 considered in a separate publication.} 
\item The particle spectrum is healthy (in particular, does not contain ghosts).
\end{enumerate} 
We note that Weyl invariance does not allow for terms quadratic in $T^A_
{\m\n}$ since they would have to come with a dimensional coefficient, and the
same goes for contributions linear in curvature $F^{AB}_{\m\n}$. As criterion
1.\ explicitly excludes Weyl-invariant terms formed from quartic powers of
$T^A_{\m\n}$, we are left with curvature-squared terms as sole ingredient for
our theory.

As is well known, curvature-squared terms generically lead to new propagating
degrees of freedom which may be ghosts (and/or tachyons of spin-1 and higher)
and are thus plagued by inconsistencies~\cite
{Stelle:1977ry,Neville:1978bk,Neville:1979rb,Sezgin:1979zf,Hayashi:1979wj,
Hayashi:1980qp,Karananas:2014pxa},  even in the absence of higher
derivatives (as is the case here). This is because the Poincar\'e group is
not compact, meaning that not all kinetic and mass terms appear with the
correct signs in the action. While unwanted particles can be removed from the
linearized spectrum by specific parameter choices~\cite
{Sezgin:1981xs,Blagojevic:1983zz,Blagojevic:1986dm,Kuhfuss:1986rb,Yo:1999ex,Yo:2001sy,Blagojevic:2002,Puetzfeld:2004yg,Yo:2006qs,Shie:2008ms,Nair:2008yh,Nikiforova:2009qr,Chen:2009at,Ni:2009fg,Baekler:2010fr,Ho:2011qn,Ho:2011xf,Ong:2013qja,Puetzfeld:2014sja,Karananas:2014pxa,Ni:2015poa,Ho:2015ulu,Karananas:2016ltn,Obukhov:2017pxa,Blagojevic:2017ssv,Blagojevic:2018dpz,Tseng:2018feo,Lin:2018awc,BeltranJimenez:2019acz,Zhang:2019mhd,Aoki:2019rvi,Zhang:2019xek,Jimenez:2019qjc,Lin:2019ugq,Percacci:2019hxn,Barker:2020gcp,BeltranJimenez:2020sqf,MaldonadoTorralba:2020mbh,Barker:2021oez,Marzo:2021esg,Marzo:2021iok,delaCruzDombriz:2021nrg,Baldazzi:2021kaf,Annala:2022gtl,Mikura:2023ruz,Mikura:2024mji,Barker:2024ydb,Barker:2024juc},
problems almost inevitably reappear in the full theory because of
strongly-coupled modes~\cite
{Moller:1961,Pellegrini:1963,Hayashi:1967se,Cho:1975dh,
Hayashi:1979qx,Hayashi:1979qx,Dimakis:1989az,Dimakis:1989ba,Lemke:1990su,Hecht:1990wn,Hecht:1991jh,Yo:2001sy,Afshordi:2006ad,Magueijo:2008sx,Charmousis:2008ce,Charmousis:2009tc,Papazoglou:2009fj,Baumann:2011dt,Baumann:2011dt,DAmico:2011eto,Gumrukcuoglu:2012aa,Wang:2017brl,Mazuet:2017rgq,BeltranJimenez:2020lee,JimenezCano:2021rlu,Barker:2022kdk,Delhom:2022vae,Annala:2022gtl,Barker:2022kdk,Barker:2023fem}.

We shall therefore focus on the only known curvature-squared terms that lead
to a healthy particle spectrum for arbitrary parameter choices.\footnote
{In general, there are 10 quadratic invariants which can be constructed from
the curvature. They are listed in~\cite{Diakonov:2011fs}.} These are composed
of the scalar curvature $F$ and its parity-odd counterpart $\tilde F$---the
so-called Holst curvature~\cite
{Hojman:1980kv,Nelson:1980ph,Castellani:1991et,Holst:1995pc}---defined as 
\be
F =  \f{1}{4}\eps_{ABCD} E^{\m\n\rho\s} e_\m^A e_\n^B F^{CD}_
{\rho\s}\ ,~~~\tilde F = E^{\m\n\rho\s} F^{AB}_{\m\n} e_{\rho A}e_
{\s B} \ , 
\ee 
where the curvature tensor $F^{AB}_{\m\n}$ was defined previously in~(\ref
{eq:curvature_tensor_F}). First, we shall  consider the most vanilla,
stripped-to-its-bare-essentials, toy-model that nevertheless still captures
the main idea---this is described by the following action \be \label
{eq:full_action} S = S_{\rm gr} + S_{\rm f} \ . \ee Here, $S_{\rm gr}$ is the
purely gravitational piece, the~\emph{quadratic EC gravitational theory},
which we take to be\footnote{It should be noted that there is freedom in
choosing the overall sign of the gravitational action. We shall discuss that
in Sec.~\ref{sec:euclidean}.}
\be
\label{eq:simple_grav_action_F_curvatures}
S_{\rm gr} = \int \diff^4 x \det (e) \l[ \f{1}{f^2} F^2 + \f{1}{\tilde f^2} \tilde F^2 \r] \ ,
\ee 
with $f$ and $\tilde f$ dimensionless constants, the couplings of the gauged
Lorentz group. Both gauge couplings will eventually have to be required to be
very small: the former ($f$) controls the cosmological constant and the
latter ($\tilde f$) the mass of a pseudoscalar mode of gravitational origin.
The theory~(\ref{eq:simple_grav_action_F_curvatures}) misses the parity-odd
term $F \tilde F$, which would be admissible according to our criteria.
Including it does not change our conclusions, as long as the coefficient in
front of it is sufficiently small, \ie does not exceed $1/f^2$ or $1/\tilde
f^2$.

For what follows it is most convenient to work in terms of the metric $g_
{\m\n}$ and the affine connection $\G^\m_{~\n\rho}$. The action~(\ref
{eq:simple_grav_action_F_curvatures}) in these variables is straightforwardly
seen to read as
\be
\label{eq:simple_grav_action}
S_{\rm gr} = \int \diff^4x \sqrt{g}\l[\f{1}{f^2} R^2 + \f{1}{\tilde f^2} \tilde R ^2 \r] \ ,
\ee 
with $g=-{\rm det}(g_{\m\n})$, and 
\be
\label{eq:scalar_curvature_Holst_curvature}
R = g^{\sigma \nu} R_{~\sigma \rho \nu}^{\rho} \ ,
~~~\tilde R = E^{\mu\nu\rho\sigma}R_{\mu\nu\rho\sigma} \ ,
\ee
where $R^\rho_{~\s\m\n}$ is the (affine) curvature tensor given in~(\ref
{eq:curvature_tensor}).

Before plunging into computations, it is instructive to do a simple counting
exercise to get a grasp on the particle spectrum of~(\ref
{eq:simple_grav_action}). In four spacetime dimensions, the connection and
tetrad carry twenty-four and sixteen degrees of freedom, respectively---not
all of them are dynamical. First of all, it is obvious that the equations of
motion are at most second order in the derivatives of the fields. Combined
with the fact that~(\ref{eq:simple_grav_action}) is invariant under local
translations (four parameters), Lorentz transformations (six parameters) and
Weyl rescalings (one parameter), leaves nineteen potentially propagating
degrees of freedom, at most; keep in mind that the Poincar\'e group ``hits
twice'' \cite{Henneaux:1992ig}. Taking into account that~\emph{in this
specific theory}~sixteen of the remaining fields are not dynamical,\footnote
{These are the components of the reduced torsion tensor.} leaves in total
three propagating degrees of freedom: the two polarizations of the massless
spin-2 graviton associated with $R^2$ and one massive pseudoscalar associated
with $\tilde R^2$.\footnote{The pseudoscalar actually descends from the
temporal component of the pseudovector torsion, exactly like a pseudoscalar
descends from the temporal  component of a pseudovector dual to the
(massive) three-form.} The presence of the latter has first been pointed out
in~\cite{Hecht:1996np}(see also the recent papers~\cite
{BeltranJimenez:2019hrm,Pradisi:2022nmh,Salvio:2022suk,Gialamas:2022xtt,Mikura:2023ruz,Barker:2024dhb}).
It is important to note that adding more curvature-squared invariants in~
(\ref{eq:simple_grav_action}), in principle increases the number of
propagating modes, one exception being  a term proportional to $R\tilde
R$.~For instance, had we included the~$(\rm{Weyl~tensor})^2$ then besides the
massless graviton and pseudoscalar, the theory would now propagate five extra
degrees of freedom corresponding to a massive spin-2 field,\footnote{This is
an aftermath of the corresponding components of the reduced torsion tensor
acquiring kinetic term and therefore becoming dynamical.}~which moreover is
known to be a ghost.

For analyzing the effect of curvature-squared terms, we shall follow the
well-known approach of introducing auxiliary fields (see \cite
{Ghilencea:2018dqd,BeltranJimenez:2019hrm,Pradisi:2022nmh,Salvio:2022suk,Gialamas:2022xtt,Barker:2024dhb}).
In our case,\footnote{The auxiliary-field method may also be employed even
when other curvature invariants are present. A well-known and studied in
details example is the Einstein-Hilber action supplemented by the square of
the Weyl tensor. The dynamics of the latter is equivalently captured~\cite
{Bergshoeff:2009hq,Ferrara:2018wlb} by a rank-2 symmetric field coupled to
the Einstein tensor and with mass term of the Pauli-Fierz type. It can be
shown that the kinetic term for the tensor field appears with wrong sign,
signaling the presence of a spin-2 ghost. } we need a dilaton $\c$ carrying
dimensions of mass and a dimensionless ``axion'' $\phi$ to recast~(\ref
{eq:simple_grav_action}) in the following equivalent form 
\be
\label{eq:simple_grav_action_auxiliary_1}
S_{\rm gr} = \int \diff^4x \sqrt{g}\Bigg[\c^2 R + M_P^2\phi \tilde R - \f{f^2\c^4}{4}-\f{\tilde f^2 M_P^4\phi^2}{4} \Bigg] \ .
\ee
Note that since the assignment of mass dimensions to the fields $\c$ and
$\phi$ is arbitrary, we made this particular choice only to simplify the
computations.

Moving on, the Weyl-invariant fermionic action $S_{\rm f}$ appearing in~(\ref
{eq:full_action}) is\footnote{For simplicity we confine ourselves to one
generation. When discussing more species, however, one may or may not  assume
that gravity is generation-blind. In the latter situation, the
torsion-fermion couplings will carry a generation index.}~\cite
{Freidel:2005sn,Alexandrov:2008iy,Karananas:2021gco,Karananas:2021zkl,Rigouzzo:2023sbb}
\be
\label{eq:fermionic_action}
S_{\rm f} = \int \diff^4 x\sqrt{g} \Bigg[ \f i 2 \overline{\Psi}\g^\m D_\m\Psi  + \text{h.c.} + \l(\z^a_V V_\m +z^a_A A_\m\r)a^\m \Bigg] \ ,
\ee
where $\z^a_V,z^a_A$ are real constants, and 
\be
V_\m = \bar{\Psi} \gamma_\mu \Psi \ ,~~~A_\m = \bar{\Psi} \gamma_5 \gamma_\mu \Psi \ ,
\ee
are the vector and axial fermionic currents, respectively. To preserve Weyl
invariance, fermions here have a minimal kinetic term and do not have a
direct coupling of the torsion vector $v^\m$ to $V_\m$ and $A_\m$. We note
that this corrects~\cite{Karananas:2021gco}, where the coupling $v^\m V_\m$
had mistakenly been included. The fermionic covariant derivative $D_\m$ is
defined as
\be
\label{eq:ferm_covariant_der}
D_\m = \mathcal D_\m +\frac 1 8 \omega_\m^{~AB}\left(\g_A\g_B - \g_B\g_A\right) \ ,
\ee
with $\mathcal D_\m$  the appropriate(flat spacetime) SM covariant derivative,
the exact form of which depends on the specific nature of the fermion
(left/right handed, lepton/quark).

\section{Equivalent metric theory}
\label{sec:metric_equivalent}

The first step in elucidating the dynamics of the action $S$ consists of using
its Weyl invariance to eliminate the dilaton by taking 
\be
\label{eq:dilaton_gauge_fixing}
\c = \f{M_P}{\sqrt 2} \ ,
\ee
such that 
\be
\label{eq:simple_grav_action_auxiliary_2}
S_{\rm gr} = M_P^2\int \diff^4x\sqrt{g}\Bigg[\f{R}{2} + \phi\tilde R -\f{\tilde f^2 M_P^2\phi^2}{4} - \f{f^2M_P^2 }{16}\Bigg] \ .
\ee
In the classical theory, we get a cosmological constant which is small (by the
smallness of the gauge coupling $f$) but cannot be zero. Needless to say,
other perturbative and non-perturbative contributions to vacuum energy are
expected to arise due to quantum effects. The Lorentz-group gravitational
gauge coupling should be chosen in a way that the total value of the vacuum
energy accounting for quantum contributions coming from the Standard Model
physics coincides with the observed value. As the SM contribution is expected
to be of the order $M_W^4$ ($M_W$ is the $W$-boson mass), an estimate of $f$
would be $f\sim (M_W/M_P)^2\sim 10^{-32}\ll 1$, though no convincing
computation can be presented. 

As a side remark, we shall point out that we can equivalently interpret the
theory \eqref{eq:simple_grav_action_auxiliary_2} in the generic metric-affine
formulation of GR, in which none of the three geometric properties curvature,
torsion and non-metricity is excluded a priori. In this case, both $R$ and
$\tilde R$ have additional contribution from non-metricity (see \cite
{Rigouzzo:2022yan,Rigouzzo:2023sbb} for details), but the behavior of the
model \eqref{eq:simple_grav_action_auxiliary_2} will still be very similar
because the Holst term $\tilde R$ exhibits an extended-projective
(EP) symmetry that is generated by a pair of vectors \cite
{Barker:2024dhb}. This invariance can be used to set non-metricity to zero,
which results in an effective equivalence of metric-affine and
Einstein-Cartan gravity.\footnote
{As it stands, the theory \eqref{eq:simple_grav_action_auxiliary_2} fails to
be EP-invariant, but this can be easily remedied with a small modification:	
\begin{align}
\label{eq:simple_grav_action_auxiliary_MAG}
S_{\rm EP} = M_P^2\int \diff^4x\sqrt{g}\Bigg[\f{\mathring{R}}{2} &+ b_1 (2 v_\mu + Q_\mu +\hat{Q}_\mu) (2 v^\mu + Q^\mu +\hat{Q}^\mu) + b_2 a_\mu a^\mu \nonumber\\
&+b_3 a_\mu (2 v^\mu + Q^\mu +\hat{Q}^\mu)  + \phi\tilde R -\f{\tilde f^2 M_P^2\phi^2}{4} - \f{f^2M_P^2 }{16}\Bigg] \ ,
\end{align}
where $b_1$, $b_2$, $b_3$ are real constants and $Q_\mu = Q_{\mu\alpha}^
{\ \ \alpha}$ and $\hat{Q}_\mu = Q^\alpha_{\ \mu\alpha}$ with non-metricity
$Q_{\alpha\mu\nu} \equiv \nabla_\alpha g_{\mu\nu}$. The metric-affine
model \eqref{eq:simple_grav_action_auxiliary_MAG} is EP-invariant and hence
fully equivalent to the EC-theory \eqref
{eq:simple_grav_action_auxiliary_2} for $b_1 = -1/12$, $b_2=1/48$ and
$b_3=0$ (\cf \eq \eqref{eq:f_decomp}). What is more, \eq \eqref
{eq:simple_grav_action_auxiliary_MAG} is the most general gravitational
theory that is symmetric under EP-transformations. Hence, one could replace
Weyl- by EP-invariance as criterion for selecting consistent models with an
ALP of gravitational origin. (Then the term $R \tilde R$ would indeed be
excluded.)}

To proceed with computations, the next step is to split the connection into a
torsion-free and a torsionful part; schematically 
\be
\label{eq:connection_split}
\Gamma \sim \mathring\Gamma + v + a +\tau \ ,
\ee
where $\mathring\Gamma$ is the usual Levi-Civita connection that depends on
the (derivatives of the) metric, while $v_\m,a_\m$ and $\tau_{\m\n\rho}$ are
the torsion vector~(\ref{eq:torsion_vector}), pseudovector~(\ref
{eq:torsion_pseudovector}) and reduced tensor~(\ref
{eq:torsion_reduced_tensor}), respectively.  In turn, the curvature tensor~
(\ref{eq:curvature_tensor}) is also decomposed into its Riemannian and
post-Riemannian pieces, which translates into the following expressions for
the scalar and Holst curvatures~(\ref{eq:scalar_curvature_Holst_curvature}) 
\begin{align}
\label{eq:f_decomp}
R & = \mathring R + 2 \mathring{\nabla}_\m v^\m -\frac 2 3 v_\m v^\m + \frac{1}{24} a_\m a^\m +\frac 1 2 \tau_{\m\n\rho}\tau^{\m\n\rho}\ ,  \\
\label{eq:tilde_f_decomp}
\tilde R & = -\mathring{\nabla}_\m a^\m+\frac 2 3 a_\m v^\m +\frac 1 2 E^{\m\n\rho\sigma}\tau_{\lambda\m\n}\tau^{\lambda}_{~\rho\sigma}\ ,
\end{align}
where $\mathring R$ is the usual, metrical, Ricci scalar.

The third step is to plug~(\ref{eq:connection_split},\ref{eq:f_decomp},\ref
{eq:tilde_f_decomp}) into~(\ref{eq:simple_grav_action_auxiliary_2}) and~(\ref
{eq:fermionic_action}), to obtain (after some integrations by parts)
\begin{align}
\label{eq:action_torsion_before}
S &=  \int \diff^4x\sqrt{g}\Bigg[M_P^2\Bigg(\f{\mathring R}{2} -\f{v_\m v^\m}{3} +\f{a_\m a^\m}{48} +\f{2\phi  a_\m v^\m}{3}  +\f{\t_{\m\n\rho}}{2}\l(\f{\t^{\m\n\rho}}{2} +\phi E^{\n\rho\kappa\lambda} \t^{\mu}_{~\kappa\lambda}\r)  \nonumber\\
& \qquad\qquad\qquad\qquad\qquad - \f{\tilde f^2 M_P^2 \phi^2}{4} - \f{f^2 M_P^2 }{16} \Bigg)+a^\m J^a_\m +\frac{i}{2} \l(\overline{\Psi}\g^\m \mathring{D}_\m\Psi +\text{h.c.} \r)\Bigg] \ ,
\end{align}
where 
\begin{align}
&J^a_\m = M_P^2\p_\m \phi + \z^a_V V_\m + \z^a_A A_\m \ ,
\end{align}
and we introduced the shifted constant~\cite{Karananas:2021zkl,Rigouzzo:2023sbb} 
\be
\z^a_A = z^a_A - \f 1 8  \ .
\ee 
Inspection of~(\ref{eq:action_torsion_before}) reveals that the connection
appears algebraically in the action and can thus be integrated out. A
straightforward computation leads to the following equations of motion for
$v,a$ and $\t$ 
\begin{align}
\label{eq:torsion_eom_1}
v_\m=-\f{24\phi J^a_\m}{M_P^2\l(1+16\phi^2\r)} \ ,~~~a_\m =- \f{24 J^a_\m}{M_P^2\l(1+16\phi^2\r)} \ ,~~~\t_{\m\n\rho} = 0 \ .
\end{align}

The last step amounts to substituting~(\ref{eq:torsion_eom_1}) into the
action~(\ref{eq:action_torsion_before}) to end up with 
\begin{align}
\label{eq:action_axion_full}
S = \f 1 2 \int &\diff^4x\sqrt{g}\Bigg(M_P^2\mathring R - \f{24M_P^2}{1+16\phi^2}(\p_\m \phi)^2- \f{\tilde f^2M_P^4\phi^2}{2} - \f{f^2 M_P^4}{8}\nonumber \\
&\quad +i \l(\overline{\Psi}\g^\m \mathring{D}_\m\Psi  + \text{h.c.} \r)-\mathscr L_{\phi V}-\mathscr L_{\phi A}-\f{3}{2M_P^2}\l( \mathscr L_{VV}+\mathscr L_{AA} +\mathscr L_{VA} \r) \Bigg) \ ,
\end{align}
where
\be
\label{eq:phi_V-phi_A}
\mathscr L _ {\phi V} = \f{48\z^a_V }{1+ 16\phi^2} V^\m \p_\m\phi \ ,~~~\mathscr L _ {\phi A} = \f{48\z^a_A}{1 + 16\phi^2}A^\m\p_\m\phi  \ ,
\ee
are contact interactions between the derivative of the ALP and the axial and
vector fermionic currents, while 
\begin{align}
&\mathscr L _ {V V} =  \f{16{\z^a_V}^2}{1 + 16\phi^2} V_\m V^\m \ ,\\
&\mathscr L _ {V A} = \f{32\z^a_A\z^a_V}{1 + 16\phi^2} V_\m A^\m \ ,\\
&\mathscr L _ {A A} = \f{16{\z^a_A}^2}{1 + 16\phi^2}A_\m A^\m \ .  
\end{align}
are the usual four-fermi interactions generically present in the EC
framework.

It is clear that the theory under consideration is the usual Einsteinian
gravity with a (nonvanishing) cosmological constant---whose value depends on
the first Lorentz gauge coupling, meaning that $f\lll 1$---coupled to fermions
and an axion-like particle with mass controlled by the second Lorentz gauge
coupling $\tilde f$. In addition, the fermionic sector comprises nontrivial
four-fermi interactions as well as mixings between $\phi$ and the vector and
axial fermionic currents. It is the latter coupling that is the crux of it
all as we shall show now.

Let us mention that the ALP nature of $\phi$ descending from $\tilde R^2$ was
also pointed out in~\cite{delaCruzDombriz:2021nrg}. The authors discussed its
dynamics, including the couplings to fermionic currents and the interactions
of $\phi$ with electromagnetism and gravity due to the chiral anomaly.
However, no coupling to QCD was included and hence the strong CP problem was
not considered.

\section{A solution to the strong CP problem}
\label{sec:strong_solution}

We focus on low energies, which amounts to taking $\phi\ll 1$. From~(\ref
{eq:action_axion_full}) we obtain
\be
\label{eq:action_axion_low-energies}
S \approx -\f 1 2 \int \diff^4x\sqrt{g}\Bigg[(\p_\m a)^2 +\f{\tilde f^2 M_P^2}{48}a^2 - 4\sqrt{6}\z^a_A\f{a}{M_P} \mathring\nabla_\m A^\m \Bigg] \ ,
\ee
where we defined $a=2\sqrt 6 M_P \phi$ to have a canonically normalized
field\footnote{There is no difficulty in canonicalizing the
ALP for all field values. This is achieved by introducing
\be
\label{eq:canonical_ALP}
a = \sqrt{\f 3 2} M_P\, \text{arcsinh}\l( 4\phi\r) \ ,
\ee
to obtain 
\begin{align}
\label{eq:action_axion_canonical}
S \supset -\f 1 2 \int \diff^4x\sqrt{g}\Bigg[(\p_\m a)^2 + \f{\tilde f^2 M_P^4}{32}\sinh^2\l(\sqrt{\f 2 3}\f{a}{M_P}\r) -12 \z^a_A  \arctan\sinh\l(\sqrt{\f 2 3}\f{a}{M_P}\r)\mathring\nabla_\m A^\m \Bigg] \ .
\end{align}
In deriving this expression we used~(\ref{eq:canonical_ALP}) to write
$\mathscr L_{\phi V}$ and $\mathscr L_{\phi A}$ in terms of the canonical
field, then absorbed the coefficient functions into the four-derivative of
$a$ and finally integrated by parts; note that  we dropped the term
proportional to the divergence of the vector current since it vanishes,
$\mathring \nabla_\m V^\m=0$. It can easily be shown that for $a\ll M_P$
(corresponding of course to $\phi\ll 1$), the action~(\ref
{eq:action_axion_canonical}) boils down to~(\ref
{eq:action_axion_low-energies}), as it should.} and retained only the terms
which are relevant for what follows.

From the above we notice that the ALP has a perturbative
``gravitationally-induced'' mass 
\be
m_{a,{\rm grav}} = \f{\tilde f M_P}{4\sqrt{3}} \ ,
\ee
whose value is determined solely by $\tilde f$; in addition and most
importantly, the field couples linearly to the (anomalous) divergence of the
axial current. In curved backgrounds this reads~\cite{Alvarez-Gaume:1983ihn}
\begin{align}
\label{eq:axial_current_divergence}
\mathring\nabla_\m A^\m &= -\f{g_1^2}{16\pi^2}E^{\m\n\rho\s}B_{\m\n}B_{\rho\s} - \f{g_2^2}{16\pi^2}E^{\m\n\rho\s} {\rm Tr }\l(\mathcal G_{\m\n} \mathcal G_{\rho\s}\r) \nonumber \\
&\qquad- \f{g_3^2}{16\pi^2}E^{\m\n\rho\s} {\rm Tr }\l(G_{\m\n}G_{\rho\s}\r) - \f{1}{384\pi^2} E^{\m\n\rho\s} \mathring R_{\m\n\kappa\lambda} \mathring R_{\rho\s}^{~~\kappa\lambda} \ ,
\end{align}
with $g_1,g_2,g_3$ the electromagnetic, weak and strong couplings, whereas $B_
{\m\n},\mathcal G_{\m\n},G_{\m\n}$ the corresponding field-strength tensors. 

Let us now focus on the ALP \& QCD contributions in the effective
theory\footnote{As is evident from \eq~(\ref
{eq:axial_current_divergence}), gravity can pose a threat to the solution to
the strong CP-problem if it introduces a gravitational theta term~\cite
{Dvali:2005an,Dvali:2017mpy,Dvali:2022fdv}. However, the solution can be
provided by the SM itself~\cite{Dvali:2013cpa,Dvali:2016uhn,Dvali:2016eay} in
the form of a composite ALP. Then the physical QCD axion is an admixture of
these pseudoscalars, so gravity is equipped with a mechanism protecting it
against itself as advocated in~\cite{Karananas:2018nrj}.}
\be
S \approx -\f 1 2 \int\diff^4x\sqrt{g}\l[(\p_\m a)^2 + m_{a,{\rm grav}}^2a^2 -\f{g_3^2}{8\pi^2}\l(\bar\theta -\f{a}{f_a}\r)E^{\m\n\rho\s} {\rm Tr }\l(G_{\m\n}G_{\rho\s}\r)  \r] \ ,
\ee
where we introduced the ALP ``decay constant''\,\footnote{Interestingly, $f_a$ is non-zero even when fermions are coupled
minimally, since in this case $\z^a_A = -\f 1 8$.}
\begin{equation}
\label{eq:ALP_decay_constant}
    f_a=\f{M_P}{2\sqrt 6 \z^a_A} \ ,
\end{equation} 
and for completeness we included the physical $\bar\theta$ angle defined as
\be
\bar \theta = \theta + {\rm arg~det}(M_u M_d) \ , 
\ee
where $\theta$ is the QCD topological angle, and $M_u$ and $M_d$ are the up-
and down- type quark mass matrices. 

Non-perturbative QCD effects generate the standard potential for the ALP and
thus the effective action at small energies becomes (see \eg \cite
{DiLuzio:2020wdo}) 
\be 
S\approx -\f 1 2 \int\diff^4x\sqrt{g}\l[(\p_\m a)^2 +m_{a,
 {\rm grav}}^2a^2 +2 m_\pi^2 f_\pi^2\sqrt{1-\f{4m_u m_d}{(m_u+m_d)^2}\sin^2\l
 (\f{\bar\theta}{2} -\f{a}{f_a} \r)}~\r] \ ,
\ee
with $m_\pi,f_\pi$ the pion mass and decay constant, and $m_u,m_d$ the up and
down quark masses. As long as the perturbative mass is smaller than the
nonperturbative one induced by QCD, then the gravi-ALP solves the strong-CP
problem. 

In this particular toy-model, this translates into
\be
\f{\tilde f}{\z^a_A} \lesssim 10^{-5} \f{m_\pi f_\pi}{M_P^2}\f{\sqrt{m_u m_d}}{m_u+m_d} \sim \mathcal O \l(10^{-43}\r) \ ,
\ee
where we took into account the observational bound on the $\bar\theta$-angle
around $10^{-10}$ (see \cite{Abel:2020pzs}). Contrary to the PQ treatment,
our proposal does not require the introduction of extra fermions or
modifications to the SM; the axion is already present in the gravitational
sector of the theory. What is more, the gravitational origin of $\tilde f$
(as well as $f$)  justifies the expectation that the coupling be vanishingly
small---remember, gravity after all is the weakest force. 

Finally, we shall mention previous attempts~\cite
{Mercuri:2009zi,Lattanzi:2009mg,Castillo-Felisola:2015ema} offering a
gravitational solution to the strong CP puzzle via an axion-like particle
associated with torsion. Crudely speaking, these models correspond to
starting from the gravitational action of our toy-model written in terms of
auxiliary fields, fixing the Weyl gauge and then setting $\tilde f
=0$---strictly speaking, this limit  cannot and should not be taken. It does
not come as a surprise that the effective low-energy dynamics and conclusions
of our construction bears similarity to the ones of the aforementioned works.
However, it must be stressed that the starting points differ radically.
Contrary to our approach, where the pseudoscalar particle is in the spectrum
of the theory from the onset, in these articles it is introduced in a
somewhat mysterious manner. Their gravitational sector comprises only the
Einstein-Hilbert term, whereas the field is a Lagrange multiplier that either
enforces the conservation of the axial torsion current (inspired by the
considerations of~\cite{Duncan:1992vz}) or the vanishing of the Nieh-Yan
topological term (inspired by the results of~\cite{Chandia:1997hu} concerning
the axial anomaly in torsionful spacetimes); interestingly, it was later
understood that this is one and the same requirement~\cite
{Castillo-Felisola:2015ema}. The bottom-line is that the pseudoscalar becomes
dynamical only after torsion is integrated out, leading to a mismatch in the
dynamical degrees of freedom.

\section{Adding the Higgs field}
\label{sec:higgs_coupling}

For the purposes of illustration, it suffices to consider the SM Higgs field
$h$ in the unitary gauge and couple it to the quadratic EC gravity in a
Weyl-invariant manner. In other words, we now take 
\be
\label{eq:action_gr_Higgs_before}
S = S_{\rm gr} + S_{\rm f} + S_{\rm Higgs}
\ee
with the purely gravitational $S_{\rm gr}$ given in~(\ref
{eq:simple_grav_action}), the fermionic one given in~(\ref
{eq:fermionic_action}), and 
\begin{align}
S_{\rm Higgs} =\f 1 2 \int \diff^4x\sqrt{g}\Bigg[\x_h h^2 R &+\z_h h^2 \tilde R +  c_{aa} h^2 a_\m a^\m + c_{\tau\tau} h^2 \tau_{\m\n\rho}^2 \nonumber\\
&+ \tilde c_{\tau\tau} h^2 E^{\m\n\rho\s}\tau_{\lambda\m\n}\tau^\lambda_{~\rho\s}-\l(D_\m^W h\r)^2-\f{\lambda h^4}{2}   \Bigg] \ ,
\end{align}
is the most general action built on the basis of Weyl invariance and the
requirement of retaining terms with at most two derivatives of the fields.
Here, $\x_h,\z_h,c_{aa},c_{\tau\tau}$ and $\tilde c_{\tau\tau}$ are (real but
otherwise arbitrary) nonminimal gravi-scalar couplings. To ensure the Weyl
invariance of the Higgs's kinetic term, the torsion vector must appear with
fixed coefficient(s) in the action and can be conveniently packed with the
derivative of $h$ to define the ``Weyl-covariant derivative'' $D_\m ^ W$ see
e.g.~\cite{Karananas:2021gco}. Explicitly, 
\be
\label{eq:weyl_covD_Higgs}
D_\m^W h = \p_\m h +\f{v_\m}{3} h \ .
\ee

In principle, we could include a Yukawa interaction between the Higgs and
fermions, however this is irrelevant for what follows. Also irrelevant are
the full expressions involving the fermionic sector, which are rather long
and do not add to the discussion; we need only worry about the interaction of
the ALP with the axial fermionic current, and we shall come back to that in
the next section. For the time being we will only consider the graviscalar
sector of~(\ref{eq:action_gr_Higgs_before}).

Following more or less verbatim the steps presented in the toy-model of the
previous section, we can easily massage~(\ref
{eq:action_gr_Higgs_before}) into its metric-equivalent form and work out its
dynamics. After fixing the Weyl gauge, see~(\ref
{eq:dilaton_gauge_fixing}), and integrating out torsion we end up with the
following Jordan-frame action
\be
S_{\rm gr} + S_{\rm Higgs} = \f 1 2 \int \diff^4 x \sqrt{g} \Bigg[\l(M_P^2+\x_h h^2\r)\mathring R -\g_{ab}\p_\m \varphi_a\p^\m\varphi_b -\f{\lambda h^4}{2}-\f{\tilde f^2 M_P^4 \phi^2}{2} - \f{f^2 M_P^4}{8}  \Bigg] \ ,
\ee
where summation over all repeated indexes is understood. Here $\varphi_a=
(h,\phi)$ and $\g_{ab}=\g_{ab}(h,\phi)$ is the metric of the internal
two-dimensional field-derivative space with components
\be
\g_{hh} =\f{N}{D} \ ,~~~\g_{h\phi} = \f{48\l(3\z_h -(1+6\x_h)\phi\r)M_P^4 h}{D} \ ,~~~\g_{\phi\phi} = \f{24\l(6M_P^2+(1+6\x_h)h^2\r)M_P^4}{D}\ ,
\ee
where 
\begin{align}
&N= 6(1+16\phi^2)M_P^4+36\l(4(\z_h^2+c_{aa})-\x_h^2 -16\x_h \z_h \phi\r)M_P^2h^2\nonumber \\
&\qquad\qquad\qquad\qquad\qquad-6\x_h\l(24\z_h^2+(\x_h+24c_{aa})(1+6\x_h)\r)h^4 \ , \\
\label{eq:denominator}
&D = 6(1+16\phi^2)M_P^4 +\l(1+12\x_h+ 144c_{aa}+96\z_h \phi\r)M_P^2 h^2\nonumber\\
&\qquad\qquad\qquad\qquad\qquad\qquad+\l(24\z_h^2 +(\x_h+24c_{aa})(1+6\x_h)\r)h^4 \ . 
\end{align}

We now transform the action in the Einstein frame to eliminate the mixing
between $h$ and gravitons. This is achieved with the usual Weyl rescaling of
the metric
\be
\label{eq:weyl_rescaling}
g_{\m\n}~\mapsto~\Omega^{-2} g_{\m\n} \ ,~~~\Omega^2 = \f{M_P^2+\x_h h^2}{M_P^2} \ .
\ee
Once this is effectuated, we find
\be
\label{eq:action_higgs_Einstein}
S_{\rm gr} + S_{\rm Higgs} =\f 1 2 \int \diff^4x \sqrt{g} \Bigg[M_P^2\mathring R -\tilde\g_{ab}\p_\m \varphi_a\p^\m\varphi_b -\f{\lambda h^4}{2\Omega^4}-\f{\tilde f^2M_P^4\phi^2}{2\Omega^4}- \f{f^2 M_P^4}{8\Omega^4} \Bigg] \ ,
\ee
where the components of the transformed field space metric $\tilde\g_
{ab}$ read
\be
\tilde \g _{hh}= \f{1}{\Omega^2}\l(\g_{hh} +\f{6\x_h^2 h^2}{M_P^2\Omega^2}\r) \ ,~~~\tilde \g_{h\phi} = \f{\g_{h\phi}}{\Omega^2} \ ,~~~\tilde \g_{\phi\phi}= \f{\g_{\phi\phi}}{\Omega^2} \ .
\ee

To make our point, it suffices to consider the low-energy limit of the theory,
so we can take $h\ll M_P$ and $\phi\ll 1$ to obtain 
\begin{align}
\label{eq:action_gr_Higgs_low-energies}
S_{\rm gr} + S_{\rm Higgs} \approx \f 1 2 \int \diff^4x \sqrt{g} \Bigg[&M_P^2 \mathring R - 24 M_P^2(\p_\m \phi)^2- \f{\tilde f^2M_P^4\phi^2}{2} - \f{f^2 M_P^4}{8} \nonumber \\
&- (\p_\m h)^2 +\f{f^2\x_h M_P^2}{4}h^2 -\f{\lambda}{2}\l(1+\f{3f^2\x_h^2}{4\lambda}\r)h^4 \Bigg] + \ldots \ ,
\end{align}
where we kept only the leading terms and the ellipses stand for higher orders
in $h$ and $\phi$. The first line in the above is exactly the
gravi-pseudoscalar sector of the previous section's toy-model, see~(\ref
{eq:action_axion_full}), and thus the dynamics are identical. From the second
line of~(\ref{eq:action_gr_Higgs_low-energies}) we observe that a
tachyonic mass for the Higgs equal to 
\be
m^2_{\rm h,grav} = -\f{f^2\x_h M_P^2}{4} \ ,
\ee
has been gravitationally induced. Since the gauge coupling $f$ sets the
cosmological constant (see the discussion after eq. (3.2)), the tree value of
the Higgs boson mass for all practical purposes can be taken to be
vanishingly small for reasonable values of the nonminimal coupling $\xi_h$.
In addition, we notice that the Higgs's quartic self-coupling gets shifted by
a gravitational contribution also proportional to $f^2$, which can be safely
neglected. 

We find it remarkable that although the inclusion of the Higgs sector is
rather non-trivial, the gravitational and ALP dynamics are practically left
unaltered as compared to the toy-model. For completeness and as mentioned
earlier, we now turn to the modifications to the relevant part of the
fermionic sector and show that the gravi-axion solution to the strong CP
problem persists. 

To this end, we focus on the contact term capturing the interaction of the ALP
with the axial current. After a straightforward but cumbersome computation we
find that in the Einstein frame this reads
\be
\mathscr L_{\phi A}= 48 M_P^2 \z^a_A \f{6M_P^2+(1+6\x_h)h^2}{D} A^\m \p_\m \phi  \ ,
\ee
with $D$ defined previously in Eq.~(\ref{eq:denominator}). As usual when
Weyl-transforming in the presence of fermions, in order to make the fermionic
kinetic term canonical, we dressed $\Psi$ with the appropriate power of the
conformal factor $\Omega$ (given in~(\ref{eq:weyl_rescaling})), i.e. we
redefined 
\be
\Psi \mapsto \Omega^{\f{3}{2}}\Psi \ .
\ee

It is easy to show that in the low-energy limit, the corresponding part of the
action becomes exactly what we found in Sec.~\ref{sec:strong_solution} 
\be
S_{\phi A} = \int \diff^4x\sqrt{g}\mathscr L_{\phi A} \approx  \int \diff^4x \sqrt{g} \f{a}{f_a} \mathring \nabla_\m A^\m \ ,
\ee
where we introduced the canonical ALP $a$ and integrated by parts; $f_a$ is
the ALP decay constant~(\ref{eq:ALP_decay_constant}).

\section{Euclidean continuation and sign of the gravitational action}
\label{sec:euclidean}

Working in the EC framework has yet another advantage, in that it incorporates
a straightforward and unambiguous way to analytically continue from
Lorentzian to Euclidean spacetimes: the Lorentz-temporal components of the
tetrad and connection get multiplied by the imaginary unit while their
purely (Lorentz-)spatial ones remain intact (see also~\cite
{Wetterich:2021ywr,Wetterich:2021hru})
\be 
e_\m^0 \to -i  e_\m^0 \ ,~~~e_\m^j \to  e_\m^j \ ,
~~~\omega_\m^{0j} \to -i \omega_\m^{0j} \ ,~~~\omega_\m^{jk} \to \omega_\m^{jk} \ ,~~~j,k = 1,2,3 \ ,
\ee
somewhat resembling the Gibbons-Hawking-Perry prescription~\cite
{Gibbons:1978ac}.~Then the curvature and torsion tensors are Euclideanized as
follows
\be 
F_{\m\n}^{0j} \to -i F_{\m\n}^{0j} \ ,~~~F_{\m\n}^{jk} \to  F_
 {\m\n}^{jk} \ ,~~~T^0_{\m\n} \to -i T^0_{\m\n} \ ,~~~T^j_{\m\n}\to T^j_{\m\n} \ .
\ee 
Applying this prescription to $S_{\rm gr}$ given in~(\ref
{eq:simple_grav_action}), we find that 
\be
 S_{\rm gr} \to - i S_{\rm gr}
\ee 
meaning that the Euclidean action is 
negative 
\be
S^{\rm E}_{\rm gr}  = - \int \diff^4 x \det(e)\l[\f{1}
 {f^2}  F^2 +\f{1}{\tilde f^2}\tilde{ F}^2 \r] \ .
\ee

Interestingly, however, there is an ``ambiguity'' in choosing the overall sign
of the EC quadratic action~(\ref{eq:simple_grav_action_F_curvatures}), see a
related discussion in \cite{Alvarez-Gaume:2015rwa}. We could have equally
well taken as our starting point 
\be
S'_{\rm gr} = -S_{\rm gr} = - \int \diff^4x \det(e)\l[\f{1}{f^2}F^2 + \f{1}{\tilde f^2}\tilde F^2 \r] \ .
\ee 
The only difference as far as the classical dynamics is concerned would be
that the classical background would correspond to an anti de Sitter
spacetime, while the (tiny) gravitationally-induced masses for the ALP and
Higgs would change their signs. Since there can be other positive
contributions to the cosmological constant, and in any case the tree-level
masses of the scalars are negligibly small, this is a perfectly acceptable
choice too.  

The Euclidean action of the sign-reverted theory is now positive-definite,
opening the possibility of its consistent path-integral formulation. This
could provide an even more solid starting point for exploring the idea of
non-perturbative Higgs mass generation~\cite{Shaposhnikov:2018xkv}, as it is
based on a positive-definite Euclidean gravitational action.

\section{Conclusions and Discussions}
\label{sec:conclusions}

In our paper we demonstrated that the combination of the Standard Model with
the Weyl-invariant Einstein-Cartan gravity leads to a remarkably economical
description of all known interactions. In addition to the fields of the
Standard Model (or $\nu$MSM) and the graviton, it automatically contains just
one extra pseudoscalar degree of freedom, or in other words an axion-like
particle (ALP), with all requisite properties to solve the strong CP problem.
The smallness of the cosmological constant and the ALP mass arise from tiny
values of the dimensionless gauge couplings of Einstein-Cartan gravity.
Moreover, the tree-level values of the Higgs (and heavy neutral lepton masses
in the $\nu$MSM) are very small or vanishing, opening up the possibility of
the potential computability of these parameters from non-perturbative
effects. Finally, yet another attractive feature of our theory is that a
particular sign choice can make its Euclidean action bounded from below,
allowing a more consistent path-integral formulation. 

Of course, our study is far from being complete. From the theoretical side,
the quantum theory of the Weyl-invariant Einstein-Cartan gravity remains to
be developed, with understanding of its high-energy limit and
non-perturbative effects, which may lead to the computation of the Higgs
boson and heavy neutral lepton masses. This theory is not pertubatively
renormalizable. We may think about its ultraviolet completion along the lines
of asymptotic safety \cite{Weinberg:1980, Reuter:1996cp, Berges:2000ew},
classicalization~\cite{Dvali:2010bf,Dvali:2010jz,Dvali:2011th}, or
non-renormalizable resummation of amplitudes proposed in~\cite
{Shaposhnikov:2023hrg}. From the phenomenological side, the most interesting
questions are related to cosmology. It would be important to understand
whether inflation can take place in this theory and what would be the
predictions of observables. The presence of a new light scalar field poses
the question whether this can be a suitable dark matter candidate. We plan to
return to these problems in the future.

A few comments are now in order. The first one concerns the role of Weyl
symmetry for our findings. A step back would be replacing Weyl invariance by
a smaller symmetry---global scale invariance. In this case, the action of the
theory contains several extra terms. In addition to the axion-like and
Standard Model particles there is an  extra propagating scalar mode---an
exactly massless dilaton, being the Goldstone boson of the spontaneously
broken scale invariance. This theory enjoys all the attractive features of
the Weyl-invariant setup, including the dynamical generation of the Planck
scale, solution of the strong CP problem, etc. In spite of the presence of a
dynamical dilaton, no long-range fifth force shows up~\cite
{Blas:2011ac,Ferreira:2016kxi}. One can go even further and remove the
requirement of the global scale invariance.~The resulting action now allows
many extra terms, including those with an explicit Planck mass. The particle
spectrum contains, as in the previous case, two extra degrees of freedom of
gravitational origin. One of them is the ALP we encountered here, which can
still be invoked to solve the strong CP problem, and the other is the
so-called scalaron, well-known from the Starobinsky inflationary model~\cite
{Starobinsky:1980te}. The computability of the Higgs and heavy neutral lepton
masses is however lost, as these terms are now allowed at tree level.

As a second comment, we believe that our proposal to solve the strong CP
problem has three advantages as compared to existing ones. First of all, our
extension of the SM and GR can be viewed as minimal since no new particles
are required apart from a single real scalar degree of freedom--the ALP
itself that is in any case required for solving the strong CP-puzzle. Second,
we do not add any additional global symmetries (or equivalently tunings of
parameters), but instead ``recycle'' the observed smallness of the
cosmological constant for explaining why the perturbative Higgs and ALP are
also tiny. Finally, gravity has provided hints that an axion must exist as a
consistency requirement, so it is intriguing that we have found a
self-consistent and purely gravitational origin for such a particle.

\section*{Acknowledgements} 

We are grateful to Will Barker and Andrey Shkerin for useful comments on the
manuscript. The work of M.S. was supported in part by the Generalitat
Valenciana grant PROMETEO/2021/083. S.Z. acknowledges the support of the
Fonds de la Recherche Scientifique - FNRS.

\appendix

\section{Diagonalization of scalar sector}
\label{app:diagonalization}

In this appendix we show that not only it is possible to fully diagonalize the
scalar kinetic sector of~(\ref{eq:action_higgs_Einstein}) by getting rid of
the derivative mixing between the Higgs and ALP, but also make canonical the
kinetic term of the former. Introduce\,\footnote{The field redefinitions can
be easily inverted to yield 
\begin{align}
\label{eq:canonical_axion_higgs_2}
&\phi = \f{3}{1+6\x_h} \l(\z_h -\f{2e^{-(1+6\x_h)\f{\Phi}{M_P}}}{1-(1+6\x_h)\tanh^2\l(\f{H}{\sqrt 6 M_P}\r)} \r) \ ,~~~h = \f{\sqrt 6 M_P\tanh\l(\f{H}{\sqrt 6 M_P}\r)}{\sqrt{1-(1+6\x_h)\tanh^2\l(\f{H}{\sqrt 6 M_P}\r)}} \ .
\end{align}
}
\bea
\label{eq:canonical_axion_higgs_1}
&\Phi = \f{M_P}{1+6\x_h}\log\l(\f{6+(1+6\x_h)\f{h^2}{M_P^2}}{3\z_h-(1+6\x_h)\phi} \r) \ ,~~~H = \sqrt 6 M_P\, \arctanh \l(\f{\f{h}{M_P}}{\sqrt{6+(1+6\x_h)\f{h^2}{M_P^2}}}\r) \ ,&
\eea
to obtain
\begin{align}
S_{\rm gr}&+S_{\rm Higgs} = \f 1 2 \int\diff^4x\sqrt{g}\Bigg[M_P^2 \mathring R -(\p_\m H)^2 -\tilde\g_{\Phi\Phi}(\p_\m\Phi)^2 -V(H,\Phi) \Bigg] \ ,
\end{align}
with
\be
\tilde\g_{\Phi\Phi} = \f 3 2 \f{\cosh^4\l(\f{H}{\sqrt 6 M_P}\r)}{\cosh^2\l(\f{H}{\sqrt 6 M_P}\r)\l(\f{1-\f{\z_h}{2}e^{(1+6\x_h)\f{\Phi}{M_P}}}{1+6\x_h}\r)^2+\l(\f{e^{(1+6\x_h)\f{\Phi}{M_P}}}{24}\r)^2\l(1+144c_{aa}\sinh^2\l(\f{H}{\sqrt 6 M_P}\r)\r)} \ ,
\ee
and 
\begin{align}
\label{eq:potential_canonical_fields}
V(H,\Phi)& =  18\lambda M_P^4\sinh^4\l(\f{H}{\sqrt 6 M_P}\r)  \nonumber \\
&\qquad+\f{f^2M_P^4}{8}\l(1-6\x_h\sinh^2\l(\f{H}{\sqrt 6 M_P}\r)\r)^2\nonumber \\
&\qquad+\f{18\tilde f^2 M_P^4 e^{-2(1+6\x_h)\f{\Phi}{M_P}}}{(1+6\x_h)^2}\Bigg[\cosh^2\l(\f{H}{\sqrt 6 M_P}\r)\nonumber \\
&\qquad\qquad\qquad-\f{\z_h}{2} e^{(1+6\x_h)\f{\Phi}{M_P}}\l(1-6\x_h\sinh^2\l(\f{H}{\sqrt 6 M_P}\r)\r)\Bigg]^2\ .
\end{align}
Interestingly, the first two terms in the potential~(\ref
{eq:potential_canonical_fields}) are identical to the ones of Ref.~\cite
{Ghilencea:2021lpa}, where the Weyl-geometric cousin of our theory was
constructed.

{
\setlength\bibsep{0pt}
    \bibliographystyle{utphys}
 \bibliography{Refs}
}

\end{document}